# Chapter 1
# Quantum Weak Measurements and Cosmology

P.C.W. Davies
Beyond Center for Fundamental Concepts in Science
Arizona State University
P.O. Box 871504, Tempe, AZ 85287

**Abstract** The indeterminism of quantum mechanics generally permits the independent specification of both an initial and a final condition on the state. Quantum pre-and-post-selection of states opens up a new, experimentally testable, sector of quantum mechanics, when combined with statistical averages of identical weak measurements. In this paper I apply the theory of weak quantum measurements combined with pre-and-post-selection to cosmology. Here, pre-selection means specifying the wave function of the universe or, in a popular semi-classical approximation, the initial quantum state of a subset of quantum fields propagating in a classical background spacetime. The novel feature is post-selection: the additional specification of a condition on the quantum state in the far future. I discuss "natural" final conditions, and show how they may lead to potentially large and observable effects at the present cosmological epoch. I also discuss how pre-and-post-selected quantum fields couple to gravity via the DeWitt-Schwinger effective action prescription, in contrast to the expectation value of the stress-energy-momentum tensor, resolving a vigorous debate from the 1970s. The paper thus provides a framework for computing large-scale cosmological effects arising from this new sector of quantum mechanics. A simple experimental test is proposed.

## 1.1 Quantum state post-selection combined with weak value measurements

One of Yakir Aharonov's most significant contributions to quantum mechanics has been to uncover the hitherto neglected sector of weak measurements combined with post-selection. In this tribute to Aharonov's pioneering work, I wish to extend his ideas to cosmology, drawing inspiration from his paper with Gruss [1]. All

P.C.W. Davies
Arizona State University, e-mail: paul.davies@asu.edu

cosmological observations and measurements are necessarily weak in the quantum sense. For example, the red shift of a galaxy will be measured by observing the light of a very large number of photons emanating from a very large number of sources within the galaxy. Although the emission of a given photon by a given atom cannot be considered as a weak process as far as the atom is concerned, the use of a large ensemble of photons to measure a property of the entire galaxy does constitute a weak measurement. The quantum back-reaction of the photons on the relevant physical variable of the whole galaxy (in this case its momentum) is negligible.

Quantum weak measurements become more interesting when combined with pre- and post-selection. (For a review of this field, with references to earlier work, see [2].) In a typical laboratory quantum weak measurement experiment, a system is prepared in a well-defined initial state $|in>$ at time $t = t_{in}$, for example, by making strong (projective) measurements of some observable *A* on an ensemble and selecting those members with the desired eigenstate. The system is then allowed to evolve, and at time $t_{out}$ another strong measurement is made, of an observable *B* (*B* is not necessarily the same as A). If this procedure is repeated for a large ensemble { *E* } of identically-prepared systems, i.e. systems all of which are in the same pre-selected eigenstate at $t = t_{in}$, and all of which are allowed to evolve with identical Hamiltonians *H*, then in general the final measured states will not be the same. Rather, they will be distributed over the set of eigenvalues of B with relative probabilities assigned by the Born rule. The experimenter then has the option of post-selecting, at time $t_{out}$, a sub-ensemble { *e* } of the total ensemble for which all the members are in a particular eigenstate of *B*; call this state $|out>$. The freedom to both pre-and-post-select the state of a system is a key property of quantum mechanics, stemming from its indeterminism. (In a closed classical Hamiltonian system, the initial state plus the Hamiltonian suffices to completely determine the final state.) If, now, the experimenter carries out weak quantum measurements of some observable C on every member of the ensemble { *E* } at one or more times t in the interval $t_{in} < t < t_{out}$, then the combined results of these weak measurements for the sub-ensemble { *e* }, satisfying both the pre- and post-selection criteria, when expressed as a statistical average, is the so-called *weak value*, given by

$$w(t) = \frac{\langle out|U^{\dagger}(t-t_{out})CU(t-t_{in})|in\rangle}{\langle out|U^{\dagger}(t-t_{out})U(t-t_{in})|in\rangle} \tag{1.1}$$

where *U* is the evolution operator $e^{-iHt}$. Weak values are not eigenvalues. Rather, they may be regarded as members of a decomposition of eigenvalues, weighted by the relative probabilities of finding the out states among { *E* }, and must be under- stood always in terms of statistical averages over large sub-ensembles: individual weak measurement results have no objective physical meaning. A good way to think about Eq. (1.1) is that the initial state is evolved forward in time from $t_{in}$ to *t*, while the final state is evolved backward in time from $t_{out}$ to *t*, and *w* is evaluated at time *t* as a composite of the forward- and backward-evolving wave functions. Note that *w* in general is not a real number, but both the real and imaginary parts have physical meaning [2]. Special interest attaches to the case that the denominator is small, which can occur if

$$Re(\langle out | U^\dagger(t - t_{out}) U(t - t_{in}) | in \rangle) \ll 1 \qquad (1.2)$$

In that case, the weak values lie well outside the spectrum of eigenvalues, and can be interpreted in some cases as an amplification process.

## 1.2 The wave function at the end of the universe: post-selection in cosmology

The concept of weak values has a clear operational meaning in a laboratory context, when the intervention of the experimenter is used to select and define appropriate in and out states. In this paper I wish to consider whether weak values with post-selection are physically meaningful in a cosmological context. In section 1, I pointed out that cosmological observations necessarily involve weak measurements. Furthermore, observational values are derived by taking statistical averages over a large ensemble (e.g. one billion photons emanating from the same galaxy), thus producing a weak value in the quantum sense. Turning now to post-selection, in the cosmological case, the problem, expressed poetically, is who or what plays the role of The Great Selector? That is, how are the states |*in*> and |*out* > to be determined?

Questions of this sort are not new in quantum cosmology. There are many proposals for defining the initial quantum state of the universe, |*in*>. Most appeal in one way or another to simplicity. A much-discussed example is the Hartle-Hawking so-called no-boundary wave function [3]. Other examples are to be found within the framework of the theory of quantum fields propagating in a classical background spacetime- a semi-classical approximation to a full theory of quantum cosmology. It is customary in this theory to take the initial quantum state to be a vacuum state of one form or another, e.g. the conformal vacuum (see, for example, [4]). The appeal to initial quantum state simplicity follows from the assumption that the universe started out physically simple and has evolved to states of greater complexity over time. Although it is far from obvious that this trend is correct [5], quantum cosmology has been practiced as a theoretical discipline for three decades without a serious challenge to the assumption of a simple initial quantum state. Often, such a state is described as "natural".

So much for pre-selection: what about post-selection- the wave function at the end of the universe, so to speak? Is there a natural |*out* > state too? Here we

encounter a fundamental prejudice that runs through all of physics. Whereas it is "natural" to suppose that the universe started out in a "special" state, few physicists consider imposing a "natural" or "special" final condition on the universe. (Some exceptions are found in [6] and [7]). The time asymmetry implicit in the discrimination between initial and final cosmological states has been the basis of much debate [8, 9, 11]. Attempts have been made [10] to construct explicitly time symmetric models of quantum cosmology, but these are considered little more than curiosities. However, in the context of weak values, there is no obligation when making a final state selection to impose time symmetry: the final state can be anything at all, so long as it is not orthogonal to the initial state evolved to the future spacetime boundary. One is therefore free to explore a number of "natural" final states in addition to the mirror image of the initial state. A proposal along these lines has been presented by Aharonov and Gruss [1].

The standard (asymmetric) treatment in quantum cosmology is to simply evolve the (simple, natural) $|in>$ state into the (usually complicated) $|out>$ state with the appropriate evolution operator:

$$|out\rangle = U(t - t_{in})|in\rangle \quad (1.3)$$

If (1.3) is substituted into Eq. (1.1), then the weak value reduces to the expectation value for observable $C$ at time $t$. But suppose, within the context of the proposal to impose pre- and post-selection criteria independently, that

$$|out\rangle \neq U(t - t_{in})|in\rangle \quad (1.4)$$

then $w(t)$ will in general not coincide with $<C>$ at time $t$. Indeed, it may differ from it dramatically in the case that condition (1.2) obtains.

By way of illustration, consider the two spacetime dimensional toy model of a scalar field with mass $m$ propagating in an expanding universe with scale factor $a(t)$ and metric

$$ds^2 = dt^2 - a^2(t)dx^2 \quad (1.5)$$

Defining the conformal time $\eta$ by

$$t = \int^t dt' = \int^\eta a(\eta')d\eta' \quad (1.6)$$

and the conformal scale factor $C$ as

$$C(\eta) = a^2(\eta) \quad (1.7)$$

it follows that

$$ds^2 = a^2(\eta)(d\eta^2 - dx^2) \tag{1.8}$$

The specific choice

$$C(\eta) = A + B\tanh(\rho\eta); \qquad A, B, \rho \text{ constants} \tag{1.9}$$

$$C(\eta) \to A \pm B, \eta \to \pm\infty \tag{1.10}$$

describes an expanding universe that is asymptotically Minkowskian in both the far past and far future, but with different scale factors $A - B$ and $A + B$. The wave equation for the massive scalar field may be explicitly solved in terms of hypergeometric functions [12]. A complete set of modes is given by

$$\begin{aligned} u_k^{in}(\eta, x) &= (4\pi\omega_{in})^{-1} \exp\left(ikx - i\omega_+\eta - (i\omega_-/\rho)\ln[2\cosh(\rho\eta)]\right) \\ &\quad {}_2F_1\left(1 + (i\omega_-/\rho), i\omega_-/\rho; 1 - (i\omega_{in}/\rho); \frac{1}{2}(1 + \tanh\rho\eta)\right) \\ &\to (4\pi\omega_{in})^{-1/2} e^{ikx - i\omega_{in}\eta}, \eta \to -\infty \end{aligned} \tag{1.11}$$

These modes coincide with standard exponential field modes in the asymptotic past. Therefore one may define a vacuum state using them. These modes may be used to define the standard quantum vacuum as $t \to -\infty$. Let us choose this vacuum state to be the "in" state of the field ($|in\rangle$) and denote it by $|0_{in}\rangle$. It corresponds to a quantum state in which there are no particles present in the "in" region (i.e. in the asymptotic past).

Another complete set of field modes is given by

$$\begin{aligned} u_k^{out}(\eta, x) &= (4\pi\omega_{out})^{-1} \exp\left(ikx - i\omega_+\eta - (i\omega_-/\rho)\ln[2\cosh(\rho\eta)]\right) \\ &\quad {}_2F_1\left(1 + (i\omega_-/\rho), i\omega_-/\rho; 1 - (i\omega_{out}/\rho); \frac{1}{2}(1 + \tanh\rho\eta)\right) \\ &\to (4\pi\omega_{out})^{-1/2} e^{ikx - i\omega_{out}\eta}, \eta \to \infty \end{aligned} \tag{1.12}$$

This second set of modes coincides with standard exponential field modes in the asymptotic future, $t \to \infty$, and may be used to define a second vacuum state $|0_{out}\rangle$, corresponding to a quantum state in which there are no particles present in the asymptotic future (the out region). These two sets of modes are not the same, so

$$|0_{in}\rangle \neq |0_{out}\rangle \tag{1.13}$$

that is, the two vacuum states, $|0_{in}\rangle$ and $|0_{out}\rangle$, are inequivalent: each corresponds to a no-particle state in its respective region, as determined by the (null) response of an inertially-moving particle detector adiabatically switched on and off again in that region. If the quantum state of the universe is chosen to be the "in" vacuum $|0_{in}\rangle$, then although there will be no particles present in the "in" region, there will be a non-zero probability for particles to be present in the "out" region. Physically, (1.13) corresponds to particles being created by the expanding universe (see, for ex- ample, [4]). (Because we work in the Heisenberg representation and there are no interactions, the evolution is contained in the time-dependence of the field modes, so the operator $U$ is trivial, and will be omitted in what follows.) The spectrum of created particles may readily be evaluated using the Bogoliubov transformation between the in and out modes

$$u_k^{in}(\eta,x) = \alpha_k u_k^{out}(\eta,x) + \beta_k u_{-k}^{out*}(\eta,x) \tag{1.14}$$

$$\alpha_k = \left(\frac{\omega_{out}}{\omega_{in}}\right)^{1/2} \frac{\Gamma(1-i\omega_{in}/\rho)\Gamma(-i\omega_{out}/\rho)}{\Gamma(-i\omega_+/\rho)\Gamma(1-i\omega_+/\rho)} \tag{1.15}$$

$$\beta_k = \left(\frac{\omega_{out}}{\omega_{in}}\right)^{1/2} \frac{\Gamma(1-i\omega_{in}/\rho)\Gamma(i\omega_{out}/\rho)}{\Gamma(i\omega_-/\rho)\Gamma(1+i\omega_-/\rho)} \tag{1.16}$$

$$\alpha_{kk'} = \alpha_k \delta_{kk'}, \quad \beta_{kk'} = \beta_k \delta_{-kk'} \tag{1.17}$$

from which it follows that

$$|\alpha_k|^2 = \frac{\sinh^2(\pi\omega_+/\rho)}{\sinh(\pi\omega_{in}/\rho)\sinh(\pi\omega_{out}/\rho)} \tag{1.18}$$

$$|\beta_k|^2 = \frac{\sinh^2(\pi\omega_-/\rho)}{\sinh(\pi\omega_{in}/\rho)\sinh(\pi\omega_{out}/\rho)} \tag{1.19}$$

where the number of particles in mode $k$ in the out region is given by Eq. (1.19).

In the foregoing treatment, the state $|0_{in}\rangle$ defines the pre-selection. In the Heisenberg representation, the state remains $|0_{in}\rangle$ throughout, including in the out region $t \to \infty$. Thus, the choice of post-selected state is the same as the pre-selected state. How-ever, we may readily generalize the treatment to the case where the post-selected state differs from the pre-selected state evolved forward to the out region. A natural choice of post-selected state is $|0_{out}>$. We may then consider weak values of observables at times between in and out. For example, consider the particle number operator

$$N_i = \Sigma_i a_i^\dagger a_i , \qquad (1.20)$$

where $a_i^\dagger$ and $a_i$ are creation and annihilation operators respectively for particles in mode $i$. The weak value of this quantity is then given by

$$w_N = \langle 0_{out}|\Sigma_i a_i^\dagger(t) a_i(t)|0_{in}\rangle / \langle 0_{out}|0_{in}\rangle \qquad (1.21)$$

To give $w_N$ a well-defined meaning, the modes corresponding to the operators $a_i^\dagger$ and $a_i$ need to be specified. One way to do this is to use the procedure of Hamiltonian diagonalization to define instantaneous modes and associated creation and annihilation operators at time $t$ during the expansion [13]. The excitations of such modes are often referred to as quasiparticles. These modes will reduce to (1.11) and (1.12) in the in and out regions respectively. It therefore follows directly from Eq. (1.21) that $w_N$ will reduce to zero in both the in region (because the state is $|0_{in}>$ there) and the out region (because the state is $|0_{out}>$ there). Observers measuring the weak values of instantaneous particle numbers present at time t will thus observe the values to rise from zero, peak around the time of maximum expansion rate, and fall towards zero again at late times. Explicit expressions for the weak values may be calculated from the mode functions and Bogoliubov transformations derived by [13] using Hamiltonian diagonalization for a massive scalar field in an expanding universe.

The toy model serves to illustrate the key issue, within the theory of quantum fields propagating in a classical background spacetime, concerning weak measurements combined with post-selection in a cosmological context, namely, the choice of final state. In the example given, it seems very natural to choose the vacuum state in the out region, as well as the (different) vacuum state in the in region. In part this is due to the fact that the spacetime is asymptotically Minkowskian in both the far past and far future. In the context of the real universe, which is (presumably) not asymptotically Minkowskian, the choice is less clear. However, it is widely accepted that the very early universe was characterized by a period of inflation, when the spacetime was a very close approximation to de Sitter space, and for which a de-Sitter invariant vacuum state is a natural choice for the initial quantum state (a popular example is the so-called Bunch-Davies vacuum state; see [14]). In addition, it is now widely accepted that the universe is dominated by dark energy, and will approach a (different, slower) de Sitter like spacetime in the far future. Therefore, it is natural to choose the final state of the universe to be the corresponding de Sitter-invariant vacuum too. If the quantum field of interest is a conformally invariant free field, then the out vacuum is merely the in vacuum evolved into the out region, and quantum post-selection will be trivial, that is, weak values will coincide with expec- tation values. In general, however, the fields will not be a conformally invariant, and will involve particle-creating interactions, so the in and out vacuum states will not be equivalent. In that case quantum post-selection of a de Sitter vacuum will result in non-trivial weak values. Depending on the specifics, there could be large weak values at the present cosmological epoch (corresponding to a small denominator in Eq. (1.1)). These would show up as cosmological anomalies – what Aharonov has referred to as quantum miracles.

To explore these quantum miracles further, I shall consider the case of anomalies in the gravitational field. To investigate this, we need to consider the back-reaction of the quantum fields on the gravitational field of the universe, in the case that we impose both pre- and post-selection on the quantum state.

## 1.3 Gravitational back-reaction

In a semi-classical theory in which the background gravitational field is treated clas- sically, and the quantum state is both pre- and post-selected, the appropriate source term to place on the right hand side of the gravitational field equations will be the weak value of the stress-energy-momentum tensor $T_{\mu\nu}$. In the case we choose vac- uum states in the "in" and "out" regions, denoted in what follows by $|0_{in}>$ and $|0_{out}>$, the gravitational equation will be

$$G_{\mu\nu} + \text{higher order terms in curvature} = \langle 0_{out}|T_{\mu\nu}|0_{in}\rangle / \langle 0_{out}|0_{in}\rangle \qquad (1.22)$$

where $G_{\mu\nu}$ is the Einstein tensor and the source term on the right hand side of Eq. (1.22) is seen to be the weak value of $T_{\mu\nu}$. Equation (1.22) was originally derived by DeWitt by adapting the Schwinger effective action theory of quantum electrodynamics to the gravitational case [15]. The effective action is defined as

$$W = i \ln(\langle 0_{out}|0_{in}\rangle) \qquad (1.23)$$

from which it follows by variation of $W$ with respect to the metric tensor $g^{\mu\nu}$ that

$$\frac{2}{(-g)^{1/2}} \frac{\delta W}{\delta g^{\mu\nu}} = \frac{\langle 0_{out}|T_{\mu\nu}|0_{in}\rangle}{\langle 0_{out}|0_{in}\rangle} \qquad (1.24)$$

Applications of Eq. (1.22) were widespread in the literature in the 1970s (see, for example, [16–18]). In the case that there is pre-selection (of state $|0_{in}\rangle$) but no post-selection, the appropriate source term to use on the right hand side of the semi-classical gravitational field equations is the expectation value $\langle 0_{in}|T_{\mu\nu}|0_{in}\rangle$:

$$G_{\mu\nu} + \text{higher order terms in curvature} = \langle 0_{in}|T_{\mu\nu}|0_{in}\rangle \quad (1.25)$$

(see, for example, [4]). It is important to note that the source terms on the right hand sides of both Eqs. (1.22) and (1.25) are formally divergent and must be renormalized [4]. It is well known that the divergent terms of both expressions are identical. However, the finite residue will in general differ, and in the case that the denominator of the right hand side of Eq. (1.22) is very small, this difference could be extremely large. The upshot is that the cosmological gravitational field of a universe in which both the initial and final states are selected could be dramatically different from one in which only the initial state is selected, the difference amounting to a gravitational quantum miracle, and being perceived by an observer as a major departure in the cosmological dynamics from what might be expected on the basis of a natural initial state alone (such as the aforementioned Bunch-Davies vacuum of inflationary cosmology). On a historical note, there was much confused debate in the 1970s about which source term, (1.22) or (1.25), to use in the semi-classical gravitational field equations. Some leading researchers (for example, [16–19]) advocated what we would now term the weak value $\langle 0_{out}|T_{\mu\nu}|0_{in}\rangle/\langle 0_{out}|0_{in}\rangle$ derived from the effective action. However, the use of the expectation value $\langle 0_{in}|T_{\mu\nu}|0_{in}\rangle$ eventually prevailed. With the benefit of hindsight, we can now see that both camps were correct. When both pre- and post-selection are involved, the weak value is indeed the appropriate source term, but if there is only pre-selection then the expectation value should be used.

The (finite) difference between the weak value and the expectation value may be formally cast in terms of the Bogoliubov coefficients $\alpha$ and $\beta$ between the states $|0_{in}\rangle$ and $|0_{out}\rangle$, as follows:

$$\langle 0_{out}|T_{\mu\nu}|0_{in}\rangle/\langle 0_{out}|0_{in}\rangle - \langle 0_{in}|T_{\mu\nu}|0_{in}\rangle = -i\Sigma_{i,j}\Lambda_{ij}T_{\mu\nu}(u^*_{in,i}, u^*_{in,j}) \quad (1.26)$$

where $T_{\mu\nu}(u^*_{in,i}, u^*_{in,j})$ is the bilinear expression for the stress-energy-momentum tensor, with the field amplitudes replaced by the mode expression for the in region (e.g. Eq. (1.11)), while

$$\Lambda_{ij} = -\, i\Sigma_k \beta_{k\,j}\, \alpha^{-1}_{ik} \qquad (1.27)$$

[19]. There is an extensive literature on how to calculate both Bogoliubov transformations and renormalized expectation values $<0_{in}|T_{\mu\nu}|0_{in}>_R$ for a variety of quantum fields in a large number of cosmological models (see [4], and references contained therein). Equation (1.26) can then in principle be used to calculate the corresponding weak values. For example, for the two dimensional expanding universe model discussed in Section 2, the renormalized expectation value of the stress-energy-momentum tensor was worked out in [12]. In that example, the Bogoliubov coefficients are given by Eqs. (1.15) and (1.16). In practice, obtaining explicit expressions for renormalized stress tensors and Bologiubov coefficients is very hard, and the manipulations involved in (1.26) and (1.27) would need to be performed numerically. However, an order of magnitude estimate may be given on dimensional grounds. For example, in the model of Section 2, significant particle production will take place only for large values of $\rho$, i.e. for rapidly expanding spacetimes, as one may infer from eq. (1.19). On general grounds, one would also expect most quantum miracles to be of the same order of magnitude. In the real universe, particle production by the cosmological expansion is negligible at the current epoch (typically one particle per Hubble volume per Hubble time), so gravitational anomalies resulting from imposing a vacuum post-selection condition on an otherwise free field are likely to be many orders of magnitude below observability. However, in the very early universe, when the rate of expansion was very high, quantum post-selection might lead to significant effects. For example, the graviton background generated in the very early universe would propagate freely to $t \to \infty$, so imposing a graviton vacuum post-selection condition would produce a large change in the gravitational source term at early times. In the case of interacting fields, such as the electromagnetic field coupled to charged matter, the resulting particle production at the present epoch is very high, and although the situation is more complicated and harder to analyze, it seems likely that a vacuum condition imposed at $t \to \infty$ would produce significant effects even at the present epoch. In Section 5, I give one suggestion for how such an effect might manifest itself.

## 1.4 Quantum miracle

As a simple illustration of a gravitational quantum miracle, consider a massless scalar field propagating in Minkowski space with pre-selected state

$$|\psi_{in}\rangle = \alpha|0\rangle + \beta|1_k\rangle \qquad (1.28)$$

for some momentum **k**, and post-selected state

$$|\psi_{out}\rangle = \gamma|0\rangle + \delta|1_{\mathbf{k}}\rangle \qquad (1.29)$$

where

$$|\alpha|^2 + |\beta|^2 = |\gamma|^2 + |\delta|^2 = 1. \qquad (1.30)$$

The weak value of the total particle number operator $N$ (see Eq. (1.20)) is then readily calculated:

$$w_N \equiv \langle\psi_{out}|N|\psi_{in}\rangle/\langle\psi_{out}|\psi_{in}\rangle = \beta\delta/(\alpha\gamma+\beta\delta) \qquad (1.31)$$

It is easy to choose values of the coefficients $\alpha, \beta, \gamma, \delta$ to yield weird weak values for $N$ (quantum miracles). For example, the choice $\alpha = \sqrt{3}/2, \beta = 1/2, \gamma = 2/\sqrt{7}, \delta = -\sqrt{3/7}$ gives

$$w_N = -1. \qquad (1.32)$$

The concept of a negative particle number also arises in the so-called quantum three-box problem discussed for example, in reference [2]. It is certainly very strange, but in the present example it attains a clear physical meaning when we calculate the weak values of $T_{\mu\nu}$ for the "in" and "out" states (1.28) and (1.29):

$$\langle\psi_{out}|T_{\mu\nu}|\psi_{in}\rangle/\langle\psi_{out}|\psi_{in}\rangle = \left(\alpha\gamma\langle 0|T_{\mu\nu}|0\rangle + \beta\delta\langle 1|T_{\mu\nu}|1\rangle\right)/(\alpha\gamma+\beta\delta), \qquad (1.33)$$

The right hand side of Eq. (1.33) is divergent. Because the field is propagating in Minkowski space, the divergence is readily renormalized by subtracting <0|$T_{\mu\nu}$|0>, this being the standard vacuum expectation value for the field (which is

conventionally set to zero). The finite part is then given, for the energy density and pressure components, by

$$\langle \psi_{out} | T_0 0 | \psi_{in} \rangle_R / \langle \psi_{out} | \psi_{in} \rangle = \beta\delta\omega/(\alpha\gamma + \beta\delta) = -\omega \qquad (1.34)$$

and

$$\langle \psi_{out} | T_1 1 | \psi_{in} \rangle_R / \langle \psi_{out} | \psi_{in} \rangle = \beta\delta\mathbf{k}/(\alpha\gamma + \beta\delta) = -\mathbf{k} \qquad (1.35)$$

where $R$ denotes renormalized. The weak value for the energy is negative, as is the weak value of the pressure component, which is a measure of the momentum of the state. A negative momentum implies that if a reflecting boundary (mirror) were introduced, then (the weak value of) the recoil of the boundary would also be negative, i.e. a weak measurement of the momentum of the mirror would indicate a shift toward the source of the incoming beam rather than away from it. If Eq. (1.33) renormalized is used as the source term in the gravitational field equations, Eq. (1.22), it will exert a negative gravitational effect, so that a weak measurement of the gravitational force would show a repulsion.

A dramatic example of what one might now call a gravitational quantum miracle has been known for nearly four decades, and was discovered by Boulware [16], who calculated the vacuum energy of a massless scalar field round a black hole using the Schwinger-DeWitt "in-out" effective action formalism, so that there is no radiation emanating from the black hole. That is, the "out" state is post-selected to be a quantum vacuum. (This contrasts with Hawking's treatment [20] in which the out state corresponds to a steady flux of thermal radiation — the so-called Hawking effect.) Boulware's treatment implies that the corresponding stress tensor, $\langle 0_{out} | T_{\mu\nu} | 0_{in} \rangle / \langle 0_{out} | 0_{in} \rangle$, diverges on the black hole event horizon, producing an infinite gravitational back-reaction [4]. Thus, imposing a vacuum post-selection condition on the universe could imply a major modification to the structure of black holes.

## 1.5 Experimental cosmology

The possibility that the quantum state of the universe might be both pre- and post-selected represents a radical departure from standard theory, and one may legitimately wonder about the possibility of experimental testing. One way is to search for gravitational and non-gravitational quantum miracles in cosmological data. Another is to perform an experiment. In this section I will describe an adaptation of an experiment performed by Partridge as a test of the time-symmetric Wheeler-Feynman theory of electrodynamics [21].

Suppose it is the case, as proposed at the end of Section 2, that the end state of the universe is a de Sitter vacuum state. Let us further suppose that such a state applies to the electromagnetic field. Then, irrespective of the CMB, and of the myriad electromagnetic interactions that are currently taking place, and will subsequently take place, throughout the universe, no photon will survive into the asymptotic future, $t \to \infty$. If it were the case that cosmological material along the future light cone of the universe absorbed all emitted photons, then the future vacuum condition would have no discernible effect on electromagnetic phenomena at the current epoch. However, it is known to be the case that the density of matter in the universe is too low for all photons emitted at our current epoch to eventually be absorbed; that is, our future light cone is not a complete absorber of photons [22, 23]. If a laser beam is directed to an empty part of the night sky, away from the galactic plane, then there is not enough matter along the line of sight of the beam to absorb all the photons, a state of affairs enhanced by the apparent accelerating expansion of the universe that serves to further dilute the density of absorbing material along the future light cone. Consider, as postulated, that the final state of the universe is an electromagnetic vacuum. Propagate the modes of this field back in time from $+\infty$ to the present epoch. Some modes will encounter matter, for example, in the interstellar medium. But, given the non-opaqueness of our future light cone, a subset of modes will travel relatively undisturbed back in time to encounter the laser. (There will be no creation of photons by the expanding universe in this case on account of the fact that the electromagnetic field is conformally invariant and a Friedmann-Roberston-Walker universe is conformally flat.) Therefore, the laser will not create photons in those modes of the electromagnetic field because, by post-selection, they have photon occupation number 0. To an experimenter, it will appear that a laser directed to an empty part of the sky will suffer a lower power drain than a laser directed at an absorbing surface. The degree of reduction will depend on the opacity of the future light cone. In the ideal case of perfect transparency, the laser would emit no energy at all in the relevant direction. In practice, of course, the effect is likely to be small, owing to the spreading of the laser beam, and hard to discern, given the gross nature of assessing the power output of a laser by measuring its power input. Nevertheless, an experiment of this sort was performed by Partridge using microwaves rather than a laser, with null results.

A more refined experiment to test the post-selected vacuum hypothesis would be to use pairs of entangled photons, which I collectively label $A$ and $B$, from a continuously pumped source. Photons $B$ are directed to a counter while their entangled partners $A$, travelling in the opposite direction, are permitted to leave the apparatus undisturbed. In the first part of the experiment, the escaped photons $A$ are intercepted by an absorber (such as a black screen on the other side of the laboratory). Maintaining this configuration, a count rate is established for photons $B$ which, by reason of the entanglement, also establishes the count rate for the emission of photons $A$. Next, the screen is retracted and photons $A$ are permitted to escape into the sky, at a variety of orientations away from any obvious absorbing material (such as the dust of the Milky Way), and the count rate of photons *B* measured again. *A* test of the post-selection hypothesis is that the count rate of photons *B* in the latter configuration would be less than in the former, because the photon source cannot excite modes of the electromagnetic field that propagate undisturbed to the postulated vacuum state at $t \to \infty$. In a perfectly transparent universe with vacuum post-selection, no photons *B* would ever be detected in the latter configuration. Cosmology is an observational science, but this experiment, along with Partridge's original, constitutes a genuine exercise in experimental cosmology.

## 1.6 Acknowledgements

I have greatly benefitted from discussions with Alonso Botero, Jeff Tollaksen and Yakir Aharonov in preparing this paper. I would like to thank Katherine Lee and Saugata Chatterjee for their help reformatting and editing this paper.